# DMINR: A Tool to Support Journalists Information Verification and Exploration

Andrew MacFarlane and Stephann Makri, Glenda Cooper, Tim Atwell
City, University of London U.K.
andym@city.ac.uk

Marisela Gutierrez-Lopez
University of Bristol, U.K

Sondess Missaoui
University of York, U.K.

Colin Porlezza
USI Università della Svizzera italiana
Switzerland

## ABSTRACT

Journalists are key information workers who have specific requirements from information systems to support the verification and exploration of information. We overview the DMINR tool, that has been designed and developed to meet the needs of journalists through the examination of journalists' information behavior in a newsroom. This study informed the design and implementation of the tool. We outline our co-design process as well as the design, implementation and deployment of the tool. We report a usability test on the tool and conclude with details of how to develop the tool further.

## CCS CONCEPTS

• Human-centered computing~Interaction design~Interaction design process and methods~Participatory design • Information systems~Information retrieval~Users and interactive retrieval ~Search interfaces

## 1  Introduction

Journalists explore information spaces to generate stories and verify the information and facts for anything published. The exploration [1] and verification [2] of information are critical working practices in the field of journalism. Journalists consult multiple sources [1] and cross check the information retrieved to ensure that the information contained is accurate, and will support story veracity [2]. This activity has now moved to the digital space [3], where there are many challenges including the variety and accuracy of sources together with the vast amount of information available to source content for a story, for instance from social media. Therefore, there is significant interest in specific tools that support the exploration and verification process both in terms of structuring and organizing a large amount of available content and sources, providing journalists a single point of access for information. Such tools often provide automated support through AI and machine learning technologies to help journalists verify and organize information [4]. This includes extracting relationships between entities such as companies and people, requiring the deployment of Natural Language Processing (NLP) methods to extract facts or identify entities and their relationships. This is the focus of the DMINR (Digital Mining In News and Retrieval) project where we are aiming to develop an entity and relationship extraction mechanism using the BERT machine learning framework [5] and provide a visualization of the data found for presentation to journalists. More details of the project can be found on the project website: https://blogs.city.ac.uk/dminr/. In this report we provide an overview of the DMINR tool, by providing the journalistic context for its development (section 2), evidence collected to support the design of the tool (section 3), the UI design process and UI design (section 4), the tool implementation (section 5) and system test (section 6). We conclude with an overview on how to develop the tools further given system test results (section 6). The contribution of the work is to provide a visualization for journalists that allows: a) exploratory search through the information space to support creativity through serendipity, and; b) providing the ability to verify information for new story ideas.

## 2  Journalistic Context

The use of AI techniques to support the journalistic workflow in the production of news is increasing, and there are many tasks to which automation can be applied including information verification and exploration [4]. The potential for AI technologies to support improvements in the workflow are considerable by improving productivity, particularly with budgets in publishing being squeezed [4]. There is also the potential to cover stories (through serendipitous discovery) that would not be covered otherwise. Recent evidence shows that journalists see the potential [6] in removing menial tasks allowing them to focus on an in-depth examination of the information for their stories, but there are significant concerns about the deployment of AI technology. There is cultural resistance [4] to the adoption of such technologies and the fear that journalists could be replaced by AI tools. Further ethical issues have arisen that have deepened this resistance to adoption [7,8]. Embedding AI technologies into journalistic workflows is therefore challenging, an issue not unknown in other fields [9,10] particularly around building trust. One way to address this is to build in key journalistic values such as transparency, accountability and responsibility into the system [11] to engender trust and promote the adoption of AI technologies in journalistic workflows. This is the starting point for our work [12].

## 3. Journalist Newsroom Study

With this journalistic context in mind, we investigated the information exploration and verification behavior of journalists in UK newsrooms to establish the requirements for our tool [13]. We used ideas directed at engendering trust as a basis for our work. To this end we conducted semi-structured interviews across the UK with 14 journalists (9 male, 5 female), in 6 large newsrooms in both public and private organizations. We used the Critical Incident Technique [14] to gather details about journalistic activities, building on more recent work to examine information interactions in creativity [15] and journalism [16,17]. We examine the journalist's role, their specialism and the set of activities they used to support writing their stories. Interviews lasted on average 35 minutes, were audio recorded and took place in a meeting room, cafe or the newsroom itself. When examining stories we asked concrete questions such as: *what inspired the story?; what made it original?; how did you find (and verify) that particular piece of information.*

We then used thematic analysis [18] to identify codes and the multidisciplinary project team (the authors) undertook a workshop to identify the five core themes in journalism practices and values to address. A second stage of coding was carried out using concepts from information science and journalism to refine the codes and provide a core set of concepts to examine [19]. This was fed into an information journey model for journalists [20] with four broad categories of activity such as: recognize a need, find information, validate and interpret and use interpretation (which in itself would initiate recognition of a need). The study identified three key design recommendations to support creativity and verification:

a) Allow journalists to conduct exploratory searches to support creativity i.e., inspiring, ideating, and developing stories.
b) Allow journalists to view documents with 'background information' to interrogate information.
c) Allow journalists to gather texts, using AI to organize and contextualize information found by search.

These recommendations are based on the assumption that journalist validate and interpret the information for their stories [3]. This involves cross-referencing information from a wide variety of difference sources to assess the accuracy, authority and credibility of the information retrieved. Often these sources will be particular to the domain e.g., business, finance, politics etc. In implementation terms this requires a single point of access for search, with the ability to plugin useful sources as and when required. Journalists tended to stack up information from the sources to see what facts and data appeared across what was retrieved, allowing triangulation and help to confirm accuracy as well as to assist creativity. An example of how this can be supported is NewsHarvester [21] where text from retrieved documents can be annotated and saved for further use. This could be augmented by AI technology by automatically identifying entities of interest (e.g., events, people, companies etc.) providing information on such entities from the different sources. Such functionality would enhance collection spaces by helping journalists to identify knowledge gaps and patterns that could point in interesting directions. By 'clicking' on a given entity of interest the journalist can explore information on that entity e.g., status of event, official address of a company or even known connections to other entities. This would provide the background information for exploration and verification, with transparency of the tool (e.g., information retrieved from specific and known sources) engendering trust. We used these design recommendations as a basis for our UI design, discussed in the next section.

## 4. DMINR UI Design

### 4.1 Co-design Workshops

Four internal project co-design workshops were conducted, with a multi-disciplinary team including interaction designers, journalists and computer scientists. A co-design methodology was developed for these workshops with journalistic values [11] in mind. These workshops were useful not only for DMINR UI design, but to integrate the findings of the journalist newsroom study into the design (see section 3 above), prioritize the system functionality, and reflect on the journalistic workflows more broadly.

The **first workshop** was dedicated to understanding the work of journalists and their perceptions of AI. Additionally, we explored the challenges and opportunities of AI in newsrooms, aiming to find "journalistic solutions to AI problems". For this workshop, we created a set of cards [22] to explore the journalistic workflow and sources of information (see figure 1). As an outcome of this workshop, we looked at how the three design recommendations could be supported in the light of the journalistic workflow and developed a series of concepts to develop and test wireframes in further internal workshops. We also considered access to multiple sources. The outcomes of the first workshop are reported in [23,24].

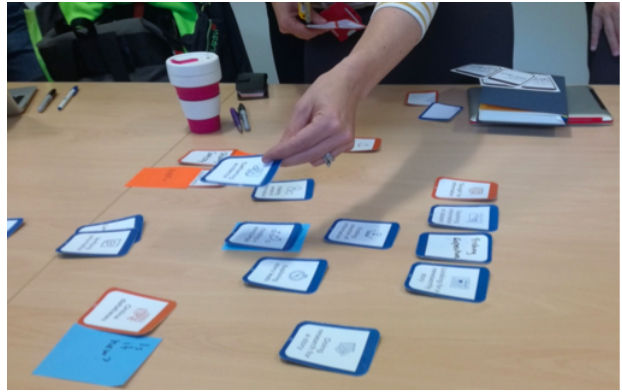

Figure 1: Card game to explore the journalistic workflow and sources of information.

The **second workshop** explored the question "what will DMINR do?", where we discussed the user needs and expectations for our tool. The discussion was facilitated by creating personas, scenarios and journey maps to identify the key touchpoints when journalists would interact with the DMINR system. To facilitate multidisciplinary discussion, we created cards to discuss the potential applications of machine learning in the context of journalistic tasks (see figure 2). As a result of this workshop, we defined and prioritized initial functionalities for the tool based on feasibility and desirability.

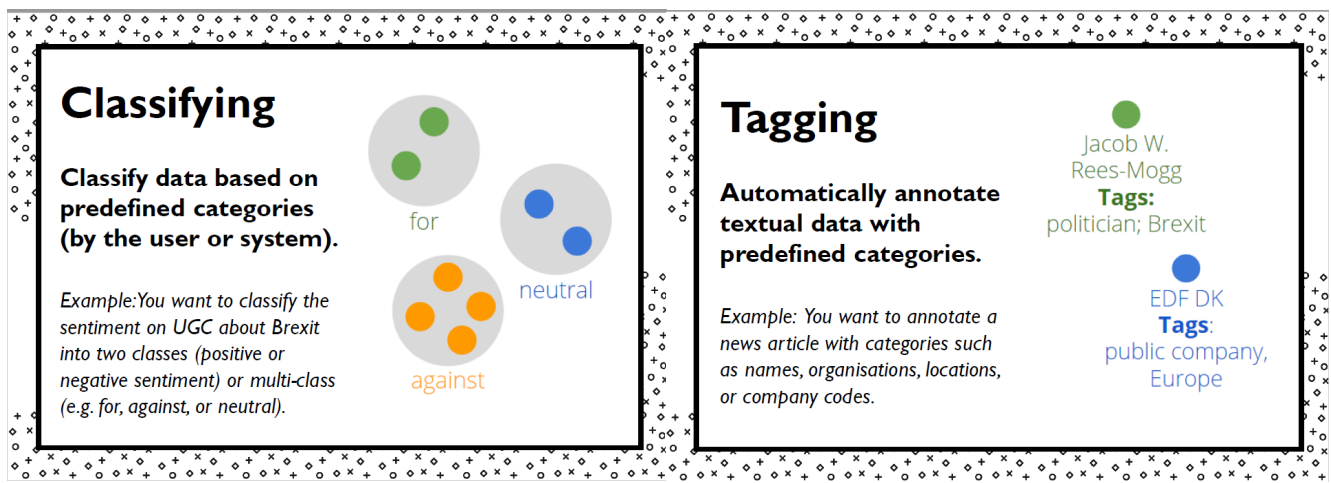

Figure 2: Examples of cards for second workshop

The **third workshop** was divided in two parts. The first part, which was facilitated by an expert UX analyst, was dedicated to creating an initial wireframe. The second part focused on refining the initial wireframe by creating a storyboard to map the available system features to journalistic workflows (e.g., which DMINR features could be useful for story discovery or verification?). This mapping allowed us to sketch design alternatives and reflect on the core interfaces required to support the journalistic workflow. The lessons learnt from this workshop helped us prioritize the implementation tasks.

The **fourth workshop** was dedicated to the design and development of the interactive visualization for DMINR. This online workshop included sketching activities using Google Slides. All workshop participants created alterative versions of the wireframe in figures 3 to 5, which helped us define the key elements for the visualization.

## 4.2 Wireframes

From the initial workshops we developed a variety of wireframes to generate ideas of what a suitable interface would look like, for supporting search and exploring entities (see figures 3 to 5). These wireframes were developed over a number of iterations, in order to produce a design that met journalists' requirements that we had found when investigating their information behavior in newsrooms. We devised a visualization that would identify the type of entity (e.g., person, company etc.) and show the connections between them, potentially labelling the type of connection e.g., through a new story or companies house record. We explored different alternatives for this visualization during the fourth co-design workshop (see above). Whilst this type of visualization is not novel in itself, it has not been developed or examined specifically for the journalism domain. This design was used as a basis for the implemented system. The visualization supports the design recommendation that arose from our newsroom investigation (see section 3):

a) Journalists are able to navigate through the entity space and explore connections to create new knowledge and verify those connections through identified sources
b) Background information on the documents that produced the entities can be provided.
c) Connections between entities are established, creating a view over the information space, providing the necessary context for creativity and verification.

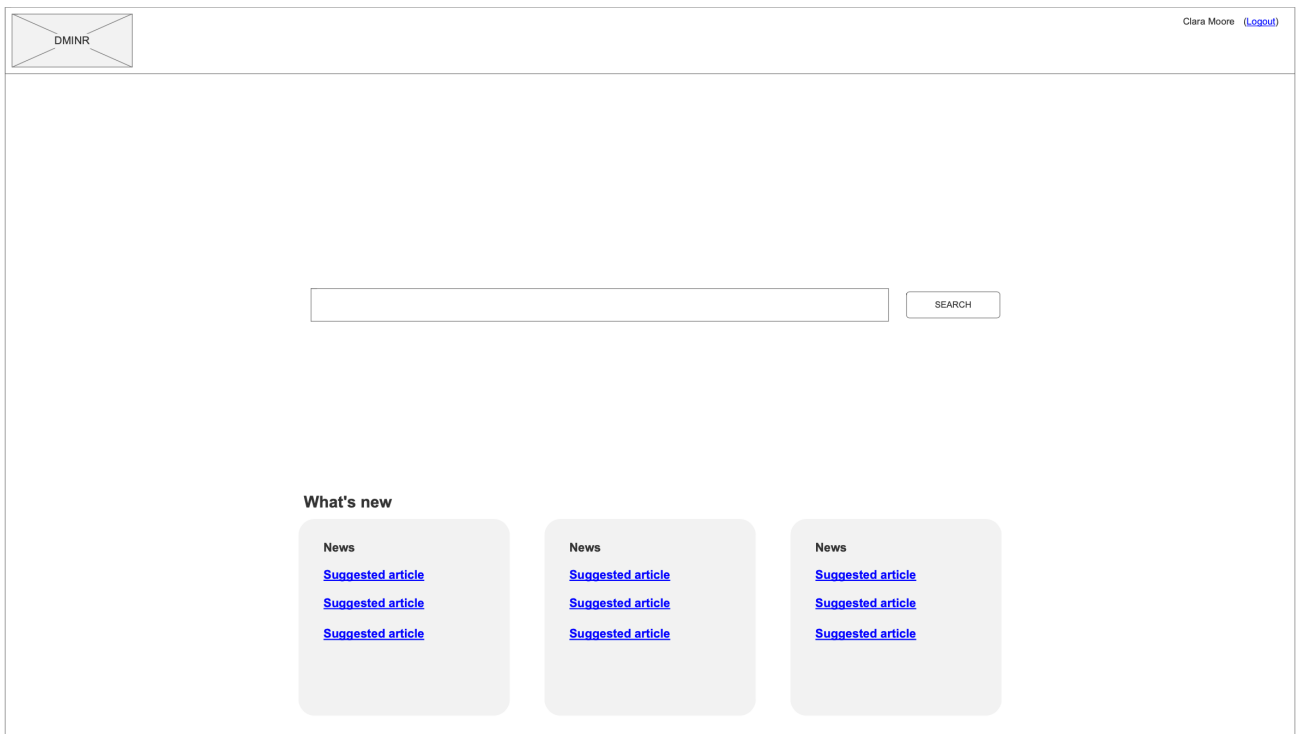

**Figure 3: Search landing page wireframe design**

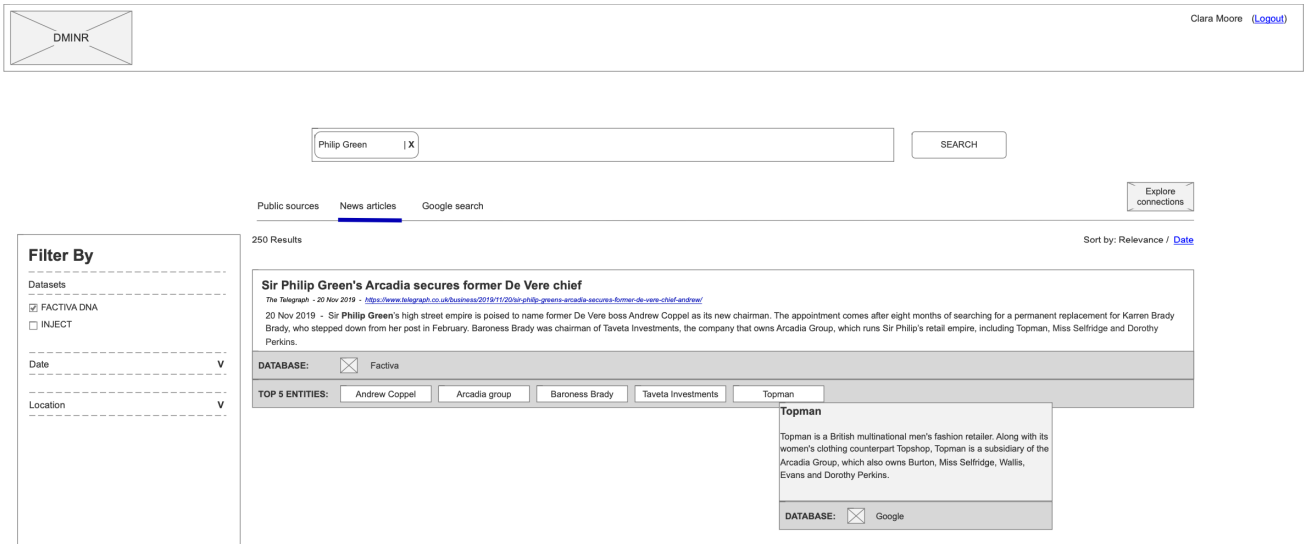

Figure 4: Results page wireframe design

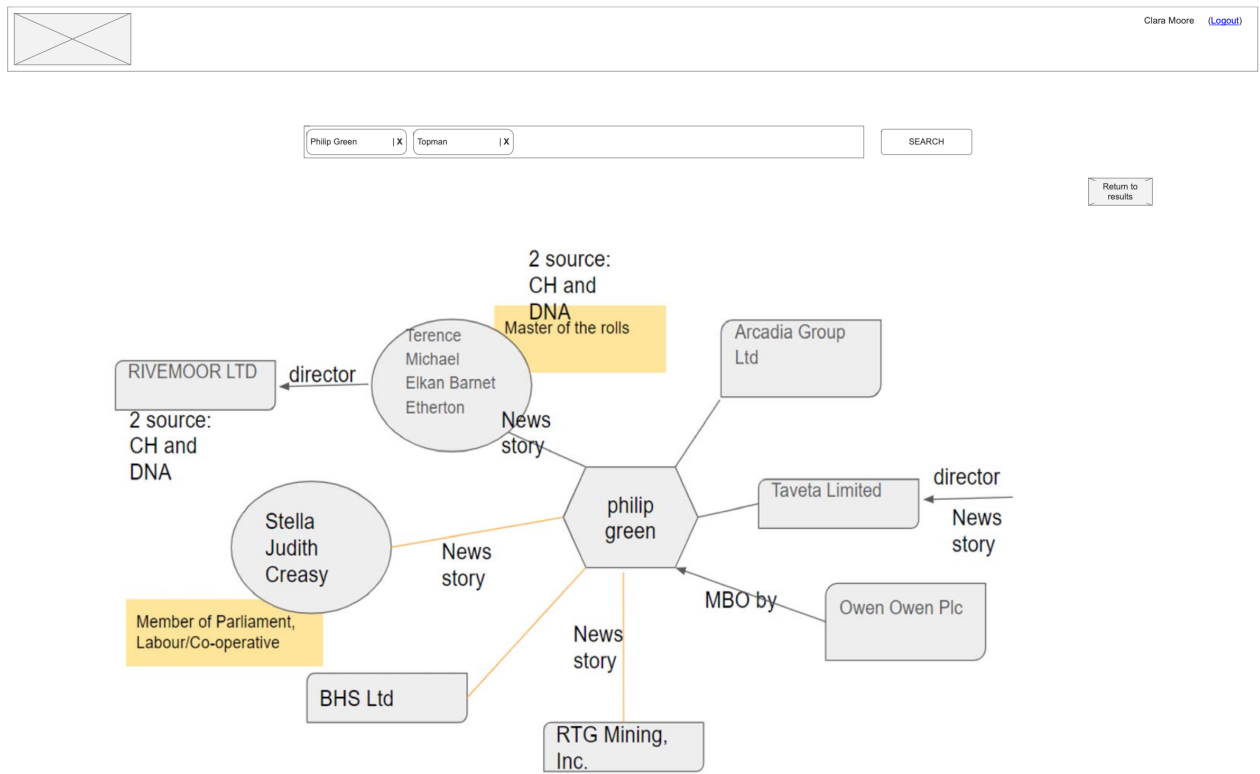

Figure 5: Exploring connections wireframe design.

## 5. DMINR System Implementation

From our design we had a number of implementation choices, such as how to extract the entities from heterogenous sources (5.1) and then display them to the user on an interface together with the basic search functionality to support the retrieval of information in order to generate the entities (5.2). Note that we did not attempt to implement all the functionality on the system due to time and resource constraints.

### 5.1 Extracting entities and establishing connections

Our requirement for extracting entities and established connections between them required the use of an NLP Technology. We considered a number of options including building our own NLP module, but after consideration we chose the BERT language model [5] as it is now widely used, very effective and allowed us to develop the software to support entity extractions very quickly. The backend system is implemented in python and PyTorch is utilized to speed up processing by parallelizing computations on a GPU. In order to support the visualizations an inverted index is created based on the user's search, and the Rust library is used to support the processing of incoming corpus by tokenizing and converting to lowercase in parallel before building the data structures for the inverted list. When testing, it was found that the scheme improved the speed of all relevant operations by at least 4, although in the data structure creation this increased to a factor of 10 or greater as the corpus from searches grew larger. An API to the entity detection module was provided for integration with the front end UI (see 5.2) below.

The process of extraction worked thus: a set of documents were retrieved from the various sources, N documents were tokenized and parsed and each token was assigned an entity label, extracting entities of particular interest for our visualization (e.g., person, company etc.). These entities were ranked using the BM25 ranking function, and the top set used for the visualization to be presented to the user. We considered a number of different BERT implementations including the original language model and DistilBERT [25]. After consideration we used DistilBERT as it was only 18% the size of the original mode and retained 97% of its accuracy, but decreased run time per query from an average of 10 seconds to 4 seconds – this represents a significant improvement in efficiency for very little loss in effectiveness.

Our entity extraction model required training, and we initially used the CONLL2003 journalistic entity extraction data set [26], but as this was somewhat old, we developed our own training set by deriving a set of topics and annotating a set of documents associated with those topics to train the BERT module. This was constructed by pulling articles a variety of reputable news sources. Articles were taken from the last seven years, with the majority being within the last three years. The project team developed a number of topics to create a general and varied corpus. Between 50 and 100 articles were used for each document and loaded in 'doccano' for labelling. The topics used were as follows: Twitter misinformation flagging, Mars Lander, Right to Repair, Aircraft, Government, Historic Scotland Building and Trade. It should be noted that the topics become increasingly general, particularly after the first three. The rationale for this is that for more niche topics, of 100 articles pulled, many would be repeats or very similar stories. This is not good for building a generalized and balanced dataset. As the search terms became more general, so did the articles they returned, and less had to be discarded.

### 5.2. DMINR UI

When developing the UI we considered [27] various aspects of the Ellis model [28] in order to think about how the interactions would work in practice. We focused on four main aspects of the model: *Monitoring*: to support access to known sources of information for the journalist; *Browsing*: through both results lists and entities; *Chaining*: examining connections between entities; and *Verification*: provide evidence through trusted sources on entities and their relationships. Exploratory browsing in the form of the Berry Picking model [29] was also supported, to allow serendipitous discovery [30] and exploratory search [31]. The front end of the system is implemented as a web application in the Django framework, to generate HTML on the fly, using a standard CCS for display consistency.

The system works as follows. Journalists can search for information and review the results on a standard Google style interface, from news sources, public sources such as Companies House or OpenCorporates and Google results (see figures 6 and 7). We provided tabs rather than federated search as Journalists will often want to consult known sources separately. The key implementation issue was the visualization which is presented in figure 8 that provided the consolidated view on all the resources. We considered a number of different types of visualizations, with the team assessing the strengths and weaknesses of each before selecting the final version shown in figure 8. We took inspiration from the Observable visualization website [32], which has many examples of visualizations in HTML to download and reuse. Two key alternatives were considered namely the use of collapsible trees which are linear in presentation [33] or force directed graphs [34]. In the end we chose the latter to allow a visualization that was both zoomable and draggable that allowed interactive exploration of the information space.

Users can explore the connections by clicking on an entity and initiating another search, and interactions to explore the graph are provided which will be useful in future developments of the tool. We produced a prototype through which our core ideas could be tested. This is the current version of the tool at the time of writing. A video of a demonstration of the tool is available [35].

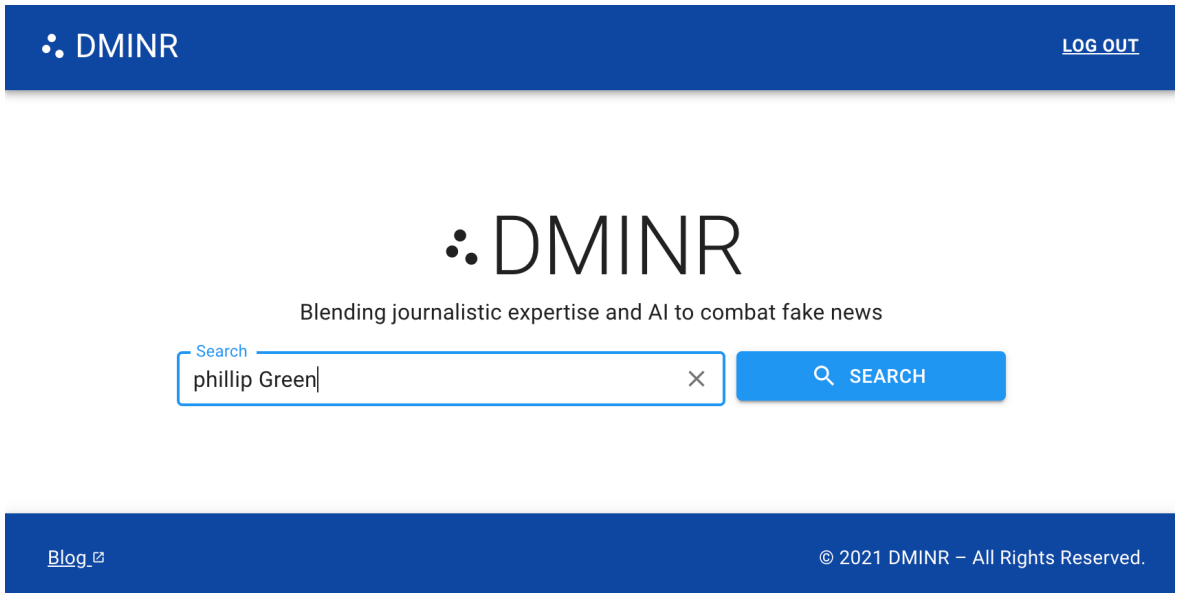

Figure 6: DMINR UI search page

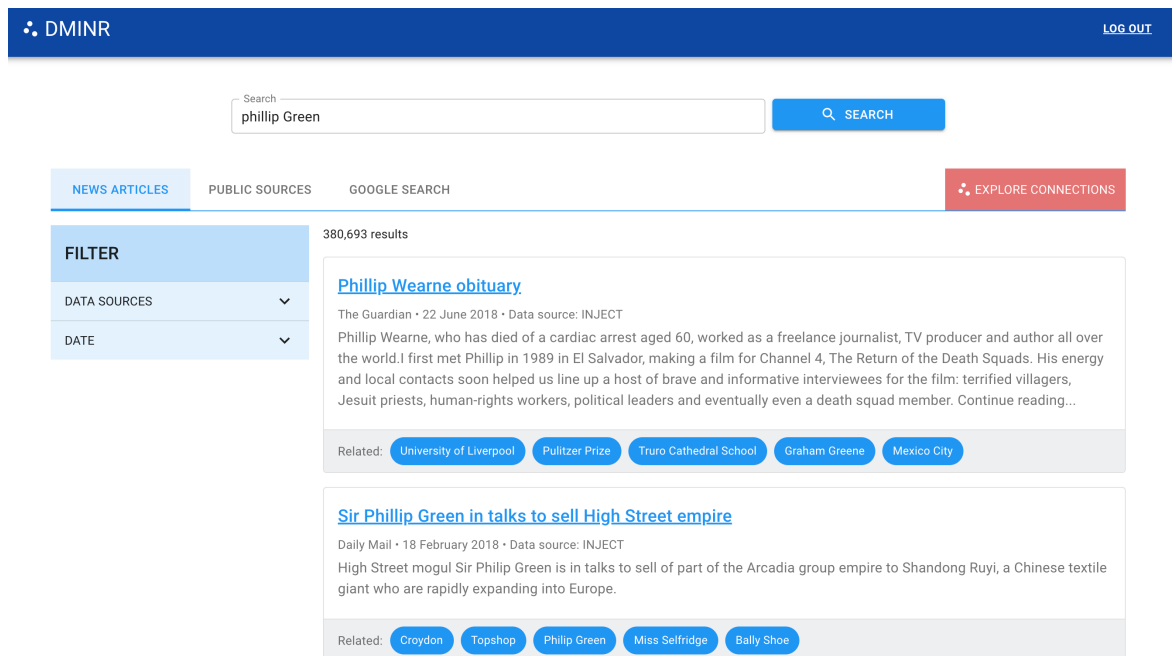

Figure 7: DMINR UI Results Tabs

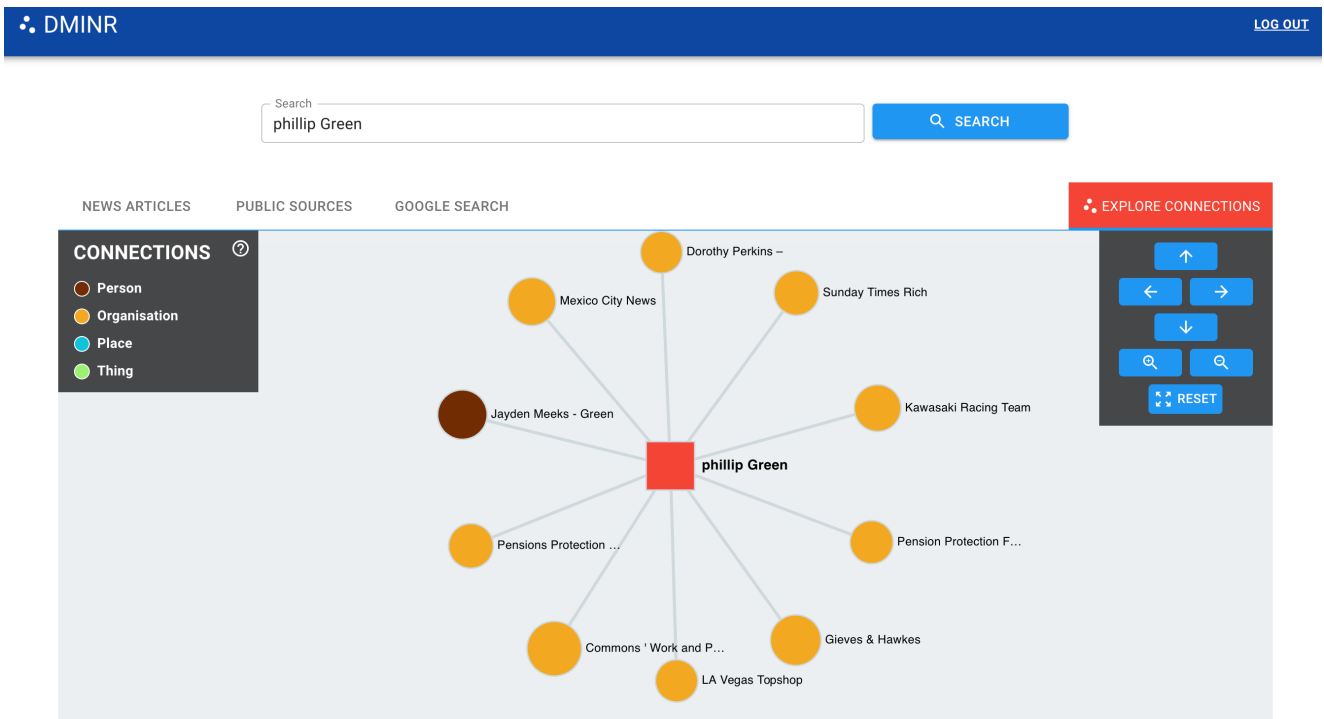

Figure 8: DMINR UI Explore Connections Tab

## 6. System Useability Test

As we did not implement all the design in our tool, we did not attempt a full scale evaluation of it. A useability test with a small number of users for our prototype was more appropriate [36]. Our test was conducted between 4[th] June 2021 and 31[st] July 2021. We recruited from two major UK news organizations. We recruited 5 users (3 Male, 2 Female) who worked as either Data Journalists or News reporters. The age range of the cohort was 27 to 36. The process worked as follows. We sent recruitment emails to our target organizations and respondents were given an introductory video to inspect, together with an information sheet and consent form. Users agreeing to the study were sent a unique username and password to access the system and given a week to undertake the study. User interactions were logged and session interactions recorded e.g. search and interaction data. On completion the users were sent a brief survey focusing on how useful the tool would be for their research and the extent to which they found the tool useful. The procedure was approved by the Departments ethics panel [37]. Users conducted on average 2.8 sessions mostly over two days, with an average query length of 2.06. Within each session all the search tabs were inspected with Article results (4.36 per session) and connections (3.21 per session dominant. Users also inspected the public sources, tending to look at information on companies (2.64 lists per session) rather than officers in those companies (0.79) per session. There were very little click throughs to retrieved items (0.64). These interactions may be a byproduct of the sample recruited, but the lack of click throughs needs to be addressed (see below). The topics used by the journalists were all news related and focused on their current role (for the purposes of privacy we do not impart these topics as they could potentially identify the journalist in question).

## 7. Conclusions and Further Work

The survey results demonstrated that users found the tools useful to some extent with 3 reporting 'slightly', 2 reporting 'moderately' and 1 reporting 'extremely'. In terms of ease of use, most users found the tool easy to use with two 'strongly agreeing and 2 'somewhat agreeing'. Only one user disagreed that the tool was useable. More positive comments included the usefulness of the connections tab and the potential for exploratory search at the beginning of a project. A number of key issues were raised by users for the tool to be addressed in further work. Access to further sources that were used in their daily work was required. It was also clear that some further work on the BERT module to detect entities and their connections was required and it did not always provide useful links. Explanations on how the tool established the connections was a concern, requiring investigation in to explainability to embed journalistic values such as transparency in the tool (raised in section 2 above).  It is also clear that the full design must be implemented in order to satisfy the often complex information seeking behaviors in exploring the information space to produce stories and verify the information for them (one user commented on the simplistic nature of the tool). Overall the tool does show some promise, but further work is required to make it useful for journalists in their story production workflow – particularly in the areas of establishing connections, explainability and comprehensiveness of the implemented design to encourage further interactions such as click throughs.

## ACKNOWLEDGMENTS

The DMINR project was funded by the Digital News Initiative. The authors are grateful to Pete Goodman and Stuart Scott for the design and implementation work they conducted to produce the DMINR tool.